\documentclass[%
 reprint,
 amsmath,amssymb,
 aps,
]{revtex4-1}

\usepackage{graphicx}
\usepackage{dcolumn}
\usepackage{bm}
\usepackage{amsmath}	
\usepackage{amssymb}	

\newcommand\PhysRevD[3]{ {Phys. Rev. D}} 
\newcommand\jcap[3]{{JCAP}} 
\newcommand\PhysRevLet[3]{ {Phys. Rev. Letters} }
\newcommand\mnras{{MNRAS}}

\usepackage[usenames]{color}
\usepackage{ulem}
\definecolor{purple}{RGB}{150,0,200}

\newcommand{\planck}{{\it Planck}}

\newcommand{\lkhd}{{\cal L}}

\newcommand{\lowz}{{\tt Low}-$z$}

\newcommand{\Om}{\Omega_{\rm{m}}}
\newcommand{\Ob}{\Omega_{\rm{b}}}
\newcommand{\ns}{n_{\rm{s}}}

\newcommand{\eone}{\Ob h^{2.55}}
\newcommand{\etwo}{\sigma_8 \Om^{0.71}}

\begin{document}


\title{Concordance Cosmology?}

\author{Park, Youngsoo}%
\affiliation{
Kavli Institute for the Physics and Mathematics of the Universe (WPI),\\
UTIAS, The University of Tokyo, Kashiwa, Chiba 277-8583, Japan}
\author{Rozo, Eduardo}
\affiliation{
Department of Physics, University of Arizona, AZ 85721, U.S.A.\\
}

\date{\today}

\begin{abstract}
We propose a new intuitive metric for evaluating the tension between two experiments, and apply it to several data sets.  While our metric is non-optimal, if evidence of tension is detected, this evidence is robust and easy to interpret. Assuming a flat $\Lambda$CDM cosmological model, we find that there is a modest $2.2\sigma$ tension between the DES Year 1 results and the {\it Planck} measurements of the Cosmic Microwave Background (CMB). 
This tension is driven by the difference between the amount of structure observed in the late-time Universe and that predicted from fitting the {\it Planck} data, and appears to be unrelated to the tension between {\it Planck} and local esitmates of the Hubble rate.
In particular, combining DES, Baryon Acoustic Oscillations (BAO), Big-Bang Nucleosynthesis (BBN), and supernovae (SNe) measurements recovers a Hubble constant and sound horizon consistent with {\it Planck}, and in tension with local distance-ladder measurements.  If the tension between these various data sets persists, it is likely that reconciling {\it all} current data will require breaking the flat $\Lambda$CDM model in at least two different ways: one involving new physics in the early Universe, and one involving new late-time Universe physics.
\end{abstract}

\pacs{Valid PACS appear here}
\maketitle


\section{Introduction}

The most recent update to the local Hubble constant measurement from \citet{Riess19} (hereafter R19) increased the tension between local measurements and the Hubble constant inferred from the \planck\ measurements \cite{Planck2018} of the Cosmic Microwave Background (CMB) to $4.4\sigma$.  In addition, multiple cosmic shear experiments appear to be in mild tension with the \planck\ results if one assumes a flat $\Lambda$CDM model \cite{des17, kidsgama2018, hsc2019}.  In a strange twist, $H_0$ estimates from combining the Dark Energy Survey (DES) Y1 results with Baryon Acoustic Oscillation (BAO) measurements \cite{bossdr12, BAO6dF, BAOSDSSMain} and Big Bang Nucleosynthesis (BBN) \cite{cookebbn} are consistent with \planck, but in tension with the R19 results.  If these tensions persist, they will signal the end of the era of ``concordance cosmology.'' 

A critical step in establishing these tensions is quantitatively evaluating the tension between various data sets under the assumption of a flat $\Lambda$CDM model in a robust fashion.  Several such methods have been proposed for in the literature, each with different strengths and weaknesses.  Here, we propose a new approach unique in that it is highly intuitive, and therefore easily lends itself to physical interpretation.  As will be shown below, our proposed tension metric points towards the key features in the flat $\Lambda$CDM model that give rise to the present day tensions.  

We propose to evaluate the tension between two experiments by identifying the directions in parameter space that are very nearly prior independent, and to use these particular sub-spaces exclusively when comparing two independent experiments.  Because the parameter combinations we identify are nearly prior independent, they can be thought of as inherent properties of the data sets themselves, making any tension discovered in these sub-spaces robust to theoretical priors.  After describing our proposed tension metric, we use it to compare a variety of data sets.  Our findings suggest that the current flat $\Lambda$CDM model fails in such a way that reconciling the model with modern cosmological data sets will require not one but two distinct modifications of the flat $\Lambda$CDM paradigm: a modification that impacts early Universe physics, and a modification that impacts late-time Universe physics (or one modification capable of affecting both the early and the late-time Universe). 

The paper is organized as follows. In Section \ref{sec:tension}, we describe our approach for quantifying tensions between constraints from independent experiments. In Sections \ref{sec:data} and \ref{sec:results}, we describe the three data sets we compare against each other (DES, \planck, \lowz, and R19) and present the results on tensions between them.  We discuss our results in Section \ref{sec:discussion}. 

\section{A Proposal for Quantifying the Tension Between Two Experiments}
\label{sec:tension}

We consider two experiments $A$ (e.g. \planck) and $B$ (e.g. DES), each of which is analyzed using the same underlying theoretical framework.  Let $p$ be the parameters shared by both experiments, and $a$ and $b$ be the parameters that apply exclusively to experiments $A$ and $B$ respectively.  We further assume that the two experiments are uncorrelated, and that experiment $A$ is more powerful than experiment $B$, as determined using the volume of the posterior over the shared parameter space.  

We assume that the likelihoods $\lkhd_A$ and $\lkhd_B$ used to analyze the experiments $A$ and $B$ are correct.  Given these likelihoods and some set of priors $P_A$ and $P_B$ (which often times are not equal to each other), each of the two data sets is analyzed via Monte Carlo Markov Chain (MCMC) realizations of the corresponding posteriors $(p_A,a)$ and $(p_B,b)$.  We assume without loss of generality that both Monte Carlo chains are of equal length. If the two chains are not of equal length, one can readily use importance sampling to generate two chains of identical length.  Given these two chains, we can construct a new chain $\Delta p = p_A - p_B$.  Note that because the experiments $A$ and $B$ are uncorrelated, and each point in the chain represent a random draw from each of the two posteriors, each element of the new $\Delta p$ chain represents a random draw of the posterior of $p_A-p_B$.  Hence, one might be tempted to establish tension between the $A$ and $B$ experiments by evaluating the probability enclosed within the posterior contour intersecting the point $\Delta p = 0$ (e.g. method 3 of \citet{charnocketal17}).

However, a problem arises when the experiments become prior dominated.  We will assume that experiment $B$, the less constraining of the two experiments, is prior dominated in some regions of parameter space.  In this case, the consistency of experiments $A$ and $B$ will depend on the priors chosen for the analysis.  It is easy to see how this dependence could lead to both artificially high or artificially low tension: overly generous priors can dilute any tension between the two experiments, whereas tight priors that are incorrectly centered could artificially increase the tension between the two experiments.

To address this difficulty we propose to focus exclusively on those directions in parameter space that are well-measured in experiment $B$.  In this context, well-measured should mean ``nearly prior independent,'' so that our conclusions about the tension between the two data sets are robust to prior assumptions. To identify the well-measured directions in parameter space, we follow the following algorithm:
\begin{enumerate}
    \item Given the MCMC samples $(p_B,b)$ of experiment B, compute the covariance matrix $C_B$ of the parameters $p_B$.
    \item Diagonalize $C_B$ to identify its eigenvectors and eigenvalues. For Gaussian distributions, these eigenvalues correspond to the variance of the corresponding eigenvectors.
    \item Sample the prior distribution for $p_B$ and project them along the measured eigenvectors. 
    \item Compare the variance obtained from the prior distribution to the variance obtained from the posterior distribution for each eigenvector of experiment B. 
    \item A particular direction in parameter space (eigenvector) is said to be well-measured if the variance of its posterior is at least 100 times smaller than the variance of its prior. 
\end{enumerate}

Of course, the threshold value of 100 is somewhat arbitrary, and an experimenter could choose to keep or throw away ``borderline'' modes for physical reasons.  So long as this decision is done {\it a priori}, the exact cutoff is irrelevant, in the sense that the resulting comparison will be a valid metric for testing the consistency of two data sets. One straightforward way to ensure the {\it a priori} nature of this decision is to select eigenmodes based on the analysis of a synthetic data set for experiment B.

\begin{figure*}[t]
\begin{center}
    \includegraphics[width=0.8\columnwidth]{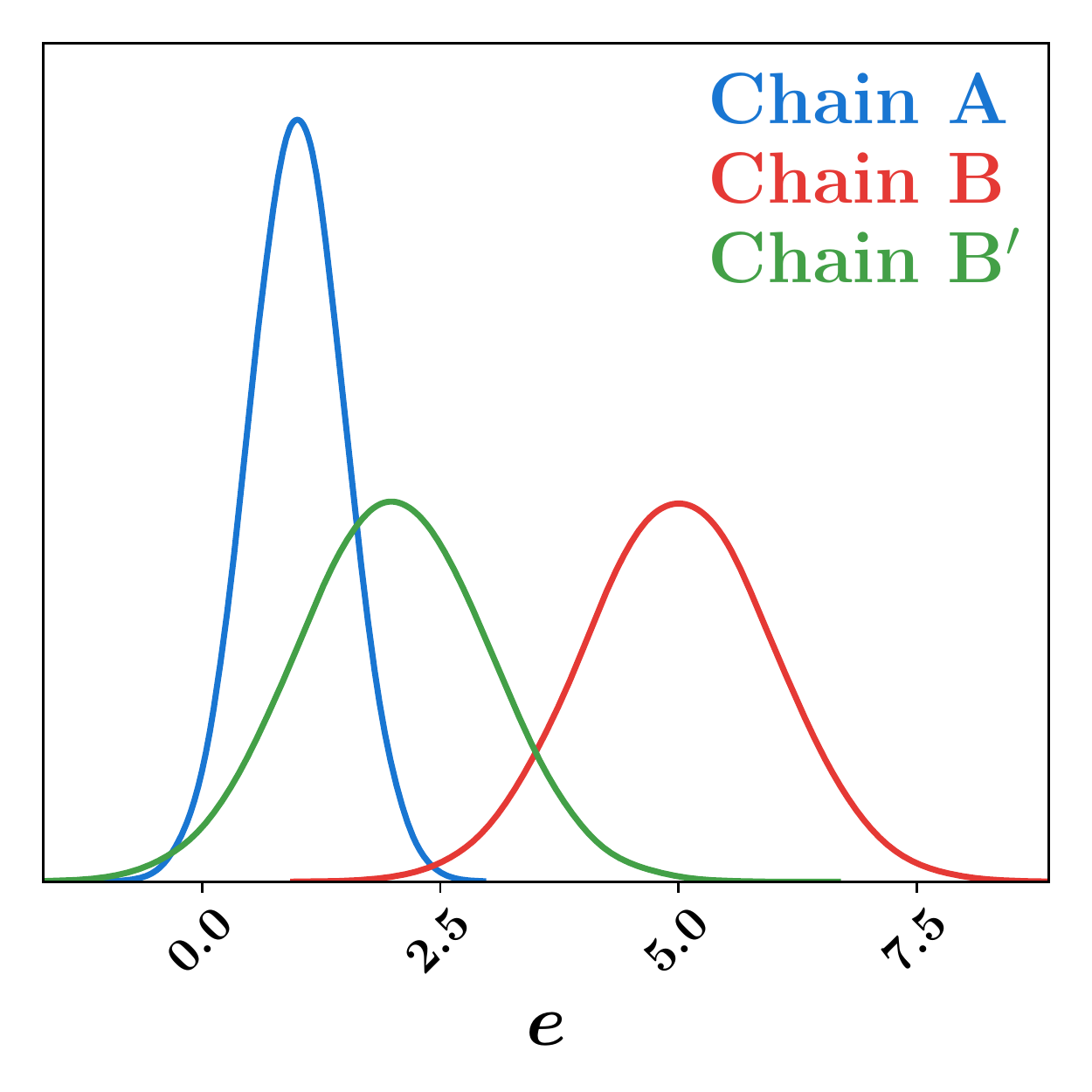}
    \hspace{0.2\columnwidth}
    \includegraphics[width=0.8\columnwidth]{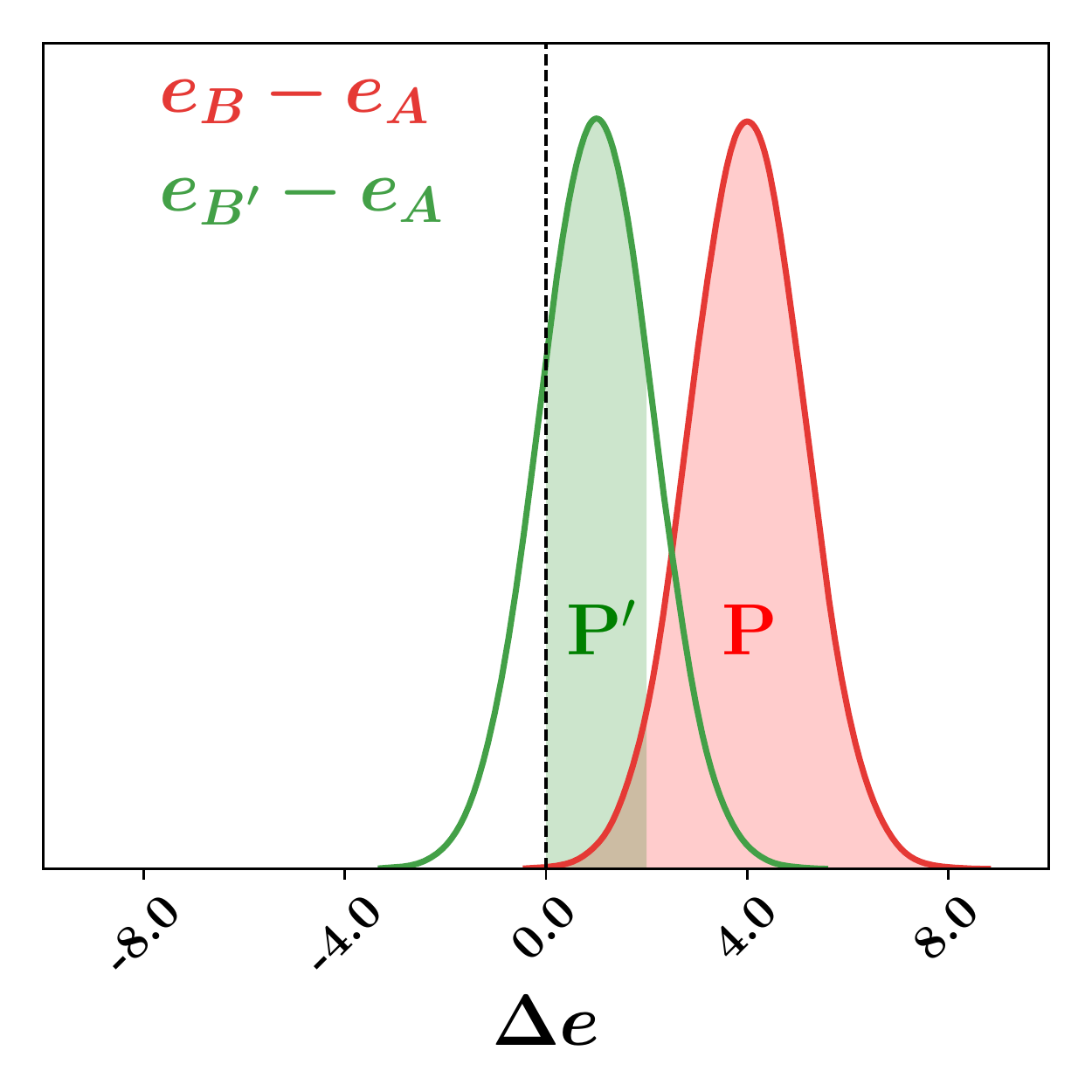}
\end{center}
\caption{{\bf Left:} Three constraints (A, B, and B') on eigenvector $e$. The consistency tests of interest are A vs. B and A vs. B'. {\bf Right:} Chains for $\Delta e$ are constructed for the consistency tests, namely $e_B-e_A$ and $e_{B'}-e_A$. From these chains, we evaluate the posterior contour that intersects $\Delta e=0$, and then integrate the posterior up to this contour.  These integrals are represented by the shaded areas P and P$^\prime$.  The recovered area is converted to a tension in Gaussian $\sigma$ via Equation \ref{eq:ptoz}.
}
\label{fig:ptoz}
\end{figure*}

To evaluate the tension between two experiments we rotate both the $p_A$ chain and the $p_B$ chain into the eigenbasis, producing two new chains $e_A$ and $e_B$.  We then generate a new chain $\Delta e$, and use this chain to evaluate the tension between the two experiments, restricting ourselves to the well-measured directions.  Specifically, we compute the contour of the posterior distribution that intersects the point $\Delta e=0$, and then evaluate the probability enclosed by this contour.  This probability can then be transformed into the usual Gaussian $\sigma$ (``$z$-$\sigma$ tension'') via the relation
\begin{equation}
P = \mbox{erf}\left( \frac{z}{\sqrt{2}}\right).
\label{eq:ptoz}
\end{equation}
A 1D example of this procedure is presented in Fig. \ref{fig:ptoz}.

We emphasize that our metric is not meant to be optimal in any way. It is simply meant to be robust to the choice of priors and easy to interpret.


\section{data sets}
\label{sec:data}

We consider four different data sets; DES, \lowz, \planck, and R19. Here, we briefly describe each of these data sets, and refer to the relevant publications for more details.

\subsection{DES}

We first consider the DES constraints obtained by combining the galaxy-galaxy \cite{desy1gc}, galaxy-shear \cite{desy1ggl}, and shear-shear \cite{desy1wl} correlation functions measured from its first-year (Y1) data set. The DES Y1 data set covers 1321 deg$^2$ in the $griz$ bands.  From this footprint, a lens sample of 650,000 redMaGiC \cite{redmagic} galaxies($0.15<z<0.9$) and a source sample containing 26 million galaxies ($0.2<z<1.3$) with shape measurements \cite{desy1shape} are identified, and used to compute galaxy--galaxy, galaxy--shear, and shear--shear angular correlation functions. Cosmological constraints from these measurements are obtained through the likelihood analysis pipeline described in \cite{desy1method}.

\subsection{\lowz}

The \lowz\ data set is comprised of the following experiments. First, we include the DES Y1 cosmology results discussed above. Second, following \cite{DESH02017, desy1ext} we add the anisotropic BAO measurements from the BOSS Data Release 12 \cite{bossdr12}. Third, we include the BBN-based measurements of $\Ob$ in \cite{cookebbn}.  In addition, as described in Section \ref{sec:results}, we verify consistency of the DES+BAO+BBN data with the {\it Pantheon} \cite{pantheon18} sample of supernovae (SNe) using our method.  Based on this agreement, we further add the {\it Pantheon} sample to our data set. We note that the BBN measurement of $\Ob$ is necessary for this data set to internally calibrate the sound horizon scale.  While this BBN constraint does not rely on low redshift Universe physics, the tensions that we discussed are not strongly impacted by the specific value of $\Ob$.  Thus, we refer to this entire data set as \lowz\ to emphasize the fact that, as we will see shortly, to the extent that this data set is in tension with \planck, this tension is driven by low redshift Universe physics. 

\subsection{\planck}
We use the final, full-mission data release from the \planck\ Collaboration \cite{Planck2018} to construct our \planck\ data set. Note that instead of the usual {\tt TTTEEE+lowl+lowE+lensing} results most frequently quoted by the \planck\ Collaboration for \planck-only cosmology constraints, we use the {\tt TTTEEE+lowl+lowE} results. This minimizes any CMB lensing information in the data set, thereby ensuring that the CMB constraints we use are driven by physics from the early (pre-recombination) Universe.

\subsection{R19}
For the distance-ladder based measurement of the Hubble constant, we use the 1D Gaussian constraint of $H_0 = 74.03 \pm 1.42\ \mathrm{km/s/Mpc}$ reported in \cite{Riess19}.

\section{Results}
\label{sec:results}


\begin{figure*}
\begin{center}
    \includegraphics[width=0.32\textwidth]{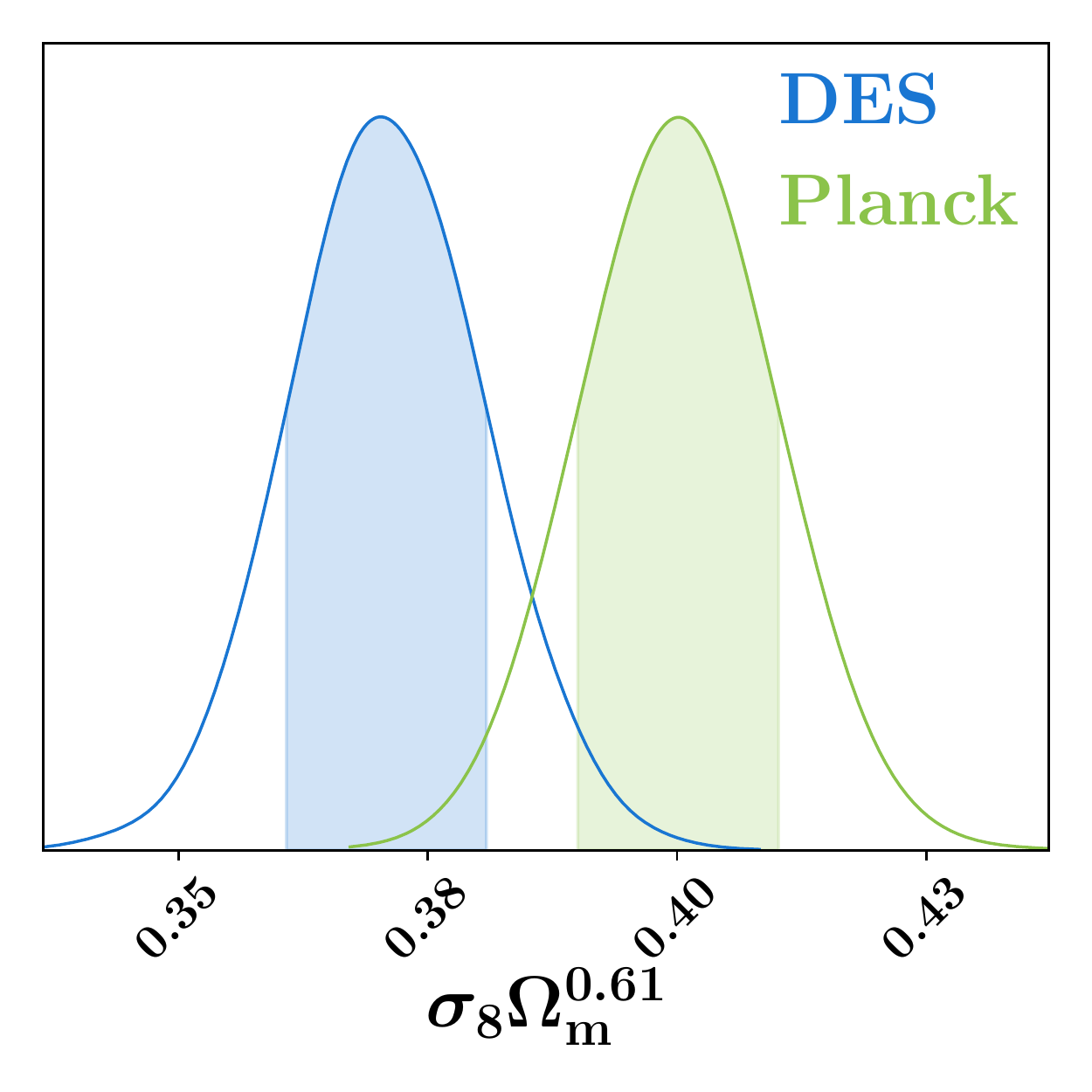}
    \includegraphics[width=0.32\textwidth]{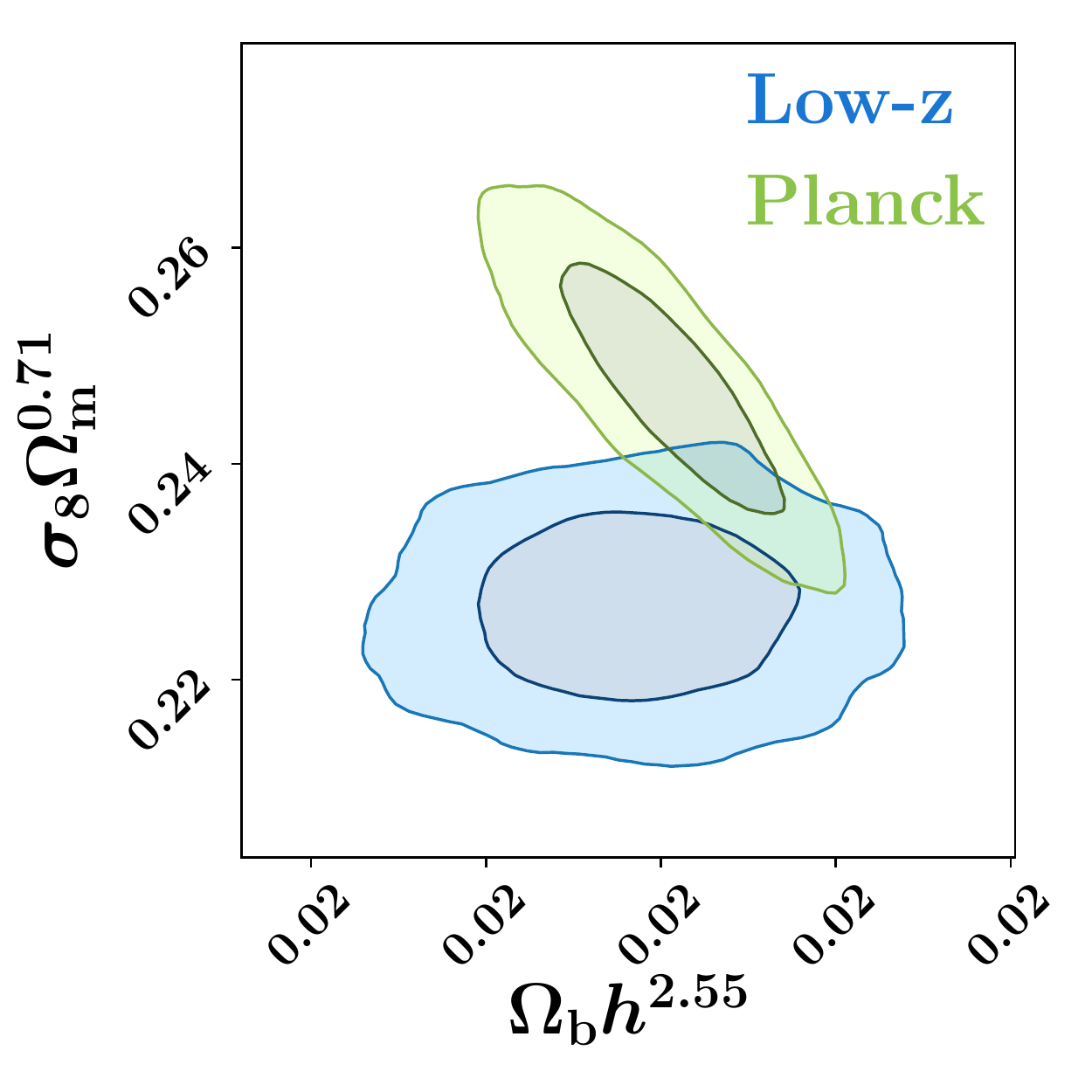}
    \includegraphics[width=0.32\textwidth]{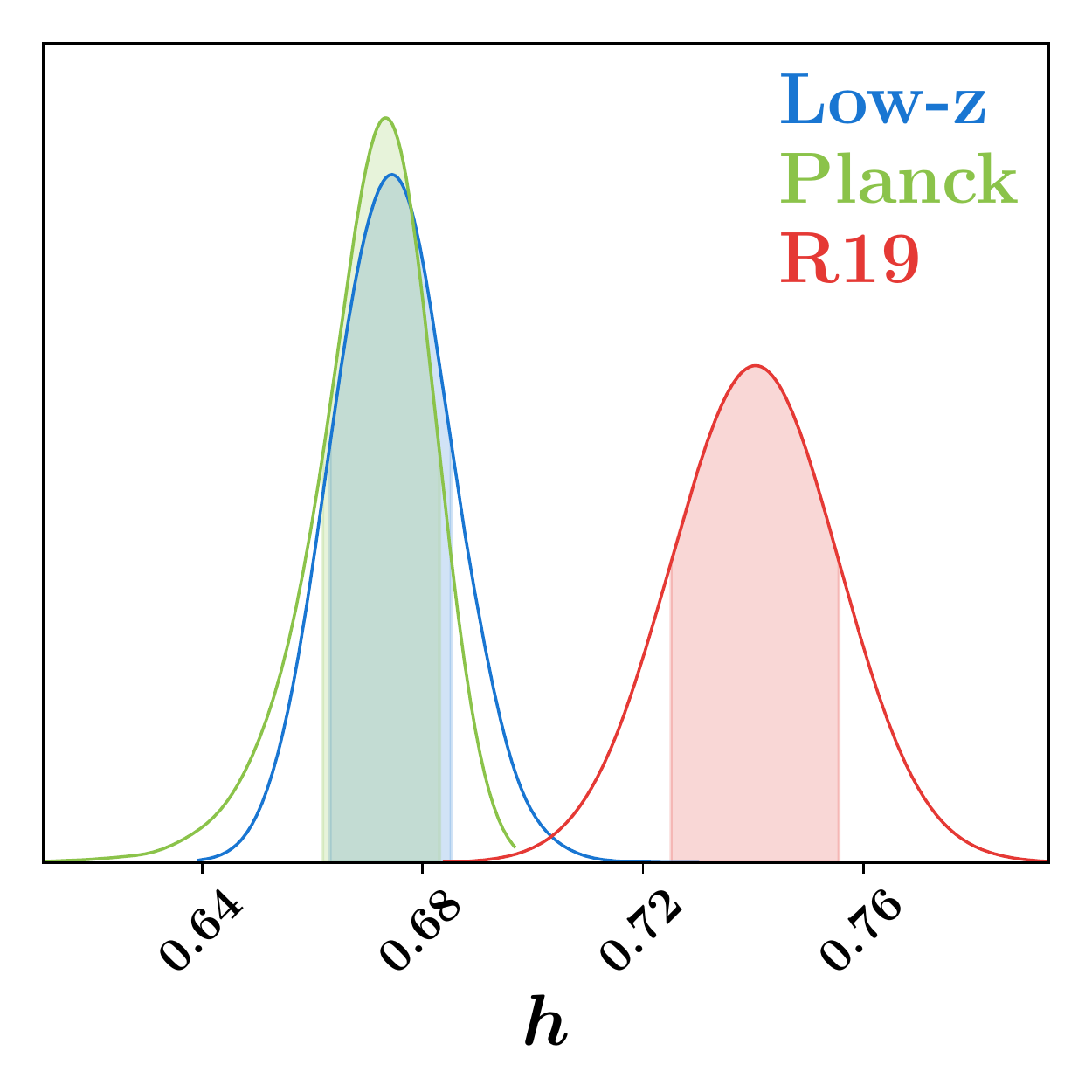}
\end{center}
\caption{{\bf Left:} Posterior distribution of the DES eigenvector $\sigma_8\Om^{0.61}$ for the \planck\ and DES data sets. {\bf Center:} Posterior distribution of the two prior-independent eigenvectors of the \lowz\ (DES+BAO+BBN+{\it Pantheon}) and \planck\ data sets.  In the plot, the eigenvector $e_1 = \eone$, while the eigenvector $e_2=\etwo$ is nearly identical to that for DES alone. {\bf Right:} Posterior distribution of the Hubble parameter $h$ for \lowz, \planck, and the R19 measurements.}
\label{fig:tension}
\end{figure*}


We begin by applying the methodology of Section~\ref{sec:tension} to evaluate the tension between the {\it Planck} and DES data sets, with \planck\ as experiment A and DES as experiment B. Throughout this section, we assume a flat $\Lambda$CDM model with free $\sum m_\nu$ and three degenerate neutrino species.  The cosmological parameters of interest are thus: $\Om$, $\Ob$, $h$, $\sigma_8$, $\ns$, and $\sum m_\nu$. In practice, rather than computing the covariance of these parameters, we evaluate the covariance matrix of the natural logarithm of these parameters.  This way, eigenvectors can be interpreted as a {\it product} of the cosmological parameters. In addition, to ensure that the robust eigenmodes are selected {\it a priori}, we employ a synthetic DES data set to generate synthetic DES posteriors. We first identify the well-measured parameter combinations from the synthetic posteriors, and then project the actual constraints along the identified robust eigenmodes. The synthetic data set for DES is constructed with the theory predictions from the DES likelihood pipeline, with fiducial parameter values listed in \cite{desy1method}.

After applying the procedure, we find that there is only one well-measured eigenvector in the DES experiment. This eigenvector is very nearly contained within the subspace $(\sigma_8,\Om)$, and is given by
\begin{equation}
    e_{\rm{DES}} \approx \sigma_8 \Om^{0.61} = 0.370 \substack{+0.011\\-0.009}.
    \label{eq:des_egvt}
\end{equation}
The corresponding value of for this same eigenvector in \citet{Planck2018} is 
\begin{equation}
    e_{\rm{Planck}} \approx 0.400 \pm 0.010.
\end{equation}
In practice, the full DES eigenvector includes very small contributions along the other parameters, but we have found that these contributions have little impact on our tension metric. Consequently, throughout this work we will restrict our presentation to these approximate eigenvectors, as they are much easier to interpret. The left panel of Figure~\ref{fig:tension} shows the posterior distribution of $e_{\rm{Planck}}$ and $ e_{\rm{DES}}$. These two measurements are at a $2.2\sigma$ tension. 

We now extend the DES data set to construct the full \lowz\ data set.
We follow \cite{DESH02017} and combine DES with the BAO and BBN data sets, and further consider adding the \textit{Pantheon} sample of SNe. To test for the consistency of the DES+BAO+BBN and \textit{Pantheon} experiments, we apply our procedure with  DES+BAO+BBN as experiment $A$ and \textit{Pantheon} as experiment $B$. In the flat $\Lambda$CDM parameter space, \textit{Pantheon} has one well-measured mode, namely
\begin{equation}
    e_{\rm{Pantheon}} = \Omega_m = 0.298 \pm 0.022,
\end{equation}
as reported in \cite{pantheon18}. The value of this same eigenvector in the DES+BAO+BBN data set is 
\begin{equation}
    \Omega_m = 0.298 \substack{+0.017\\-0.018}.
\end{equation}
The two measurements are in excellent agreement, and thus we may safely combine them. This combination yields the full \lowz\ data set, upon which we apply our procedure and find two eigenmodes that are well-measured:
\begin{eqnarray}
    e_1 & = & \eone, \\
    e_2 & = & \etwo.
\end{eqnarray}
The second eigenmode ($e_2$) is nearly identical to the original DES eigenmode, while the first eigenmode ($e_1$) represents the  constraining power on $h$ and $\Omega_\mathrm{b}$ newly gained by the combination of BAO, BBN, and SNe data sets. Note that again we are using a synthetic data set for \lowz\ in identifying $e_1$ and $e_2$. This synthetic \lowz\ data set is constructed in a similar fashion to the aforementioned synthetic DES data set, i.e. with theory predictions for the four experiments evaluated at the fiducial parameter value specified in \cite{desy1method}.


\begin{figure*}
\begin{center}
    \includegraphics[width=0.7\textwidth]{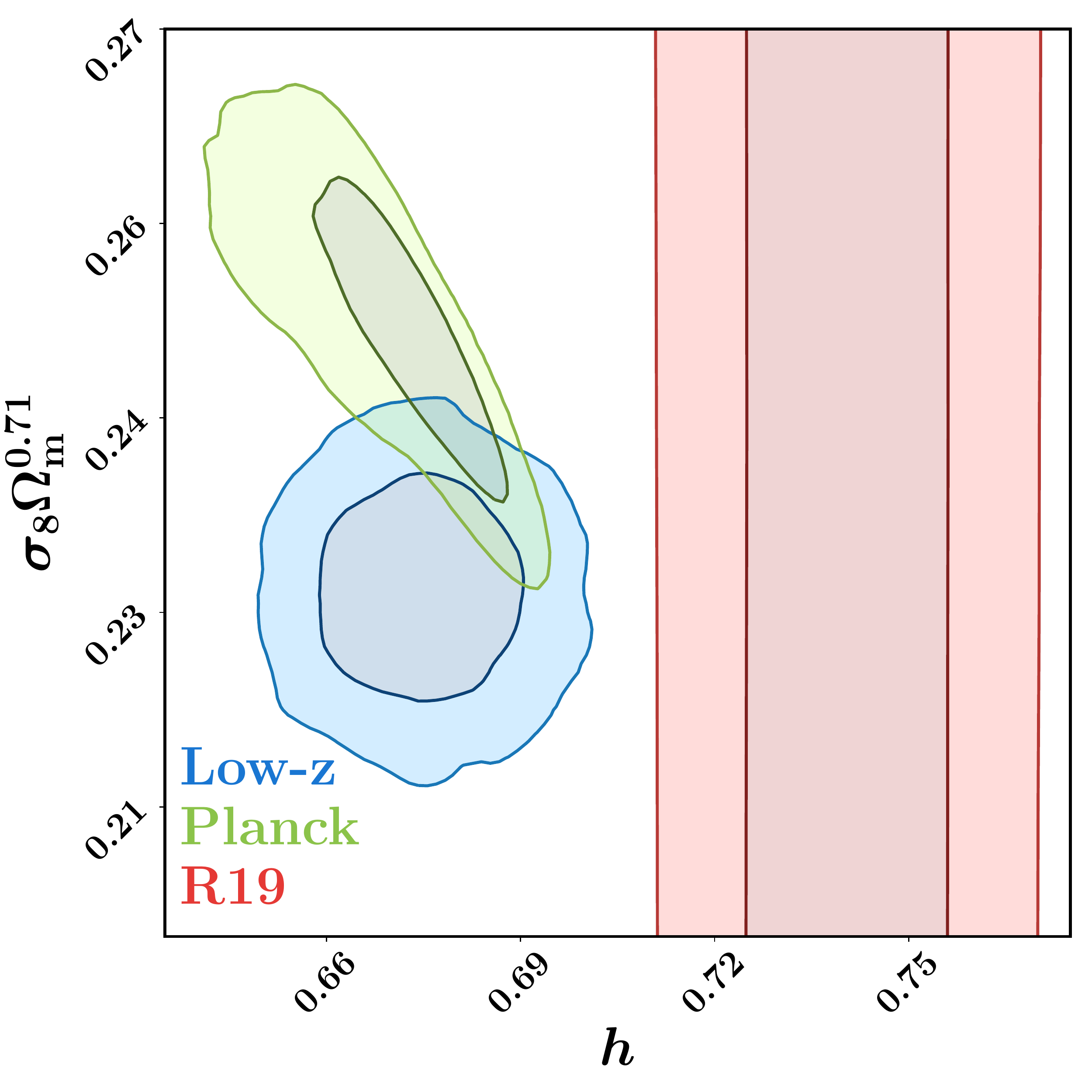}
\end{center}
\caption{\lowz, \planck, and R19 constraints in the $e_2$--$h$ ($\etwo$--$h$) parameter subspace.  We emphasize that the parameter combination $\etwo$ is very nearly prior independent in the \lowz\ data set, and is associated with the lensing strength of the low redshift Universe.  The Hubble parameter constraint from the \lowz\ data set is driven by the same early Universe physics that lead to the \planck\ counterpart, but constitutes a fully independent calibration of the sound horizon scale.  The excellent agreement between \planck\ and \lowz\ on the Hubble parameter paints a consistent picture of the sound horizon, and suggests that reconciling these values with R19 will require modification to early Universe physics. In contrast, the $e_2$ tension between \planck\ and \lowz\ data sets reflect a difference between the lensing strength predicted from \planck\ based on the early Universe and the low redshift measurements. Reconciling these two results will therefore require altering late-time Universe physics, likely those associated with the current accelerated phase of expansion.  Put together, the results shown here point towards two very distinct failures of the standard flat $\Lambda$CDM paradigm.}
\label{fig:tension2}
\end{figure*}


We are now ready to perform the three-way consistency test between \planck, \lowz, and R19, via the following experiment pairs:
\begin{itemize}
    \item \planck\ (A) vs. \lowz\ (B)
    \item \planck\ (A) vs. R19 (B)
    \item \lowz\ (A) vs. R19 (B)
\end{itemize}
The middle panel of Figure~\ref{fig:tension} shows the results from the first comparison, namely between \planck\ and \lowz. The two data sets are in good agreement along the $e_1$ direction, while exhibiting tension along the $e_2$ direction. Compared with the \planck\ vs. DES pair, the overall tension in the well-measured 2D parameter subspace is reduced to $1.9\sigma$ because of the look-elsewhere effect (if we were to only consider the $e_2$ direction, the estimated tension would have increased to $2.3\sigma$).
The right panel of Figure~\ref{fig:tension} shows the results from the second and third comparisons. As R19 plays the role of experiment B here, we show the posterior distributions of the Hubble parameter, i.e. the only eigenmode in R19, for the \planck, \lowz, and R19 data sets. We recover a $4.0\sigma$ tension between \planck\ and R19, and a $3.7\sigma$ tension between the \lowz\ and R19 data sets.

\section{Discussion}
\label{sec:discussion}


\begin{figure*}
\begin{center}
    \includegraphics[width=0.32\textwidth]{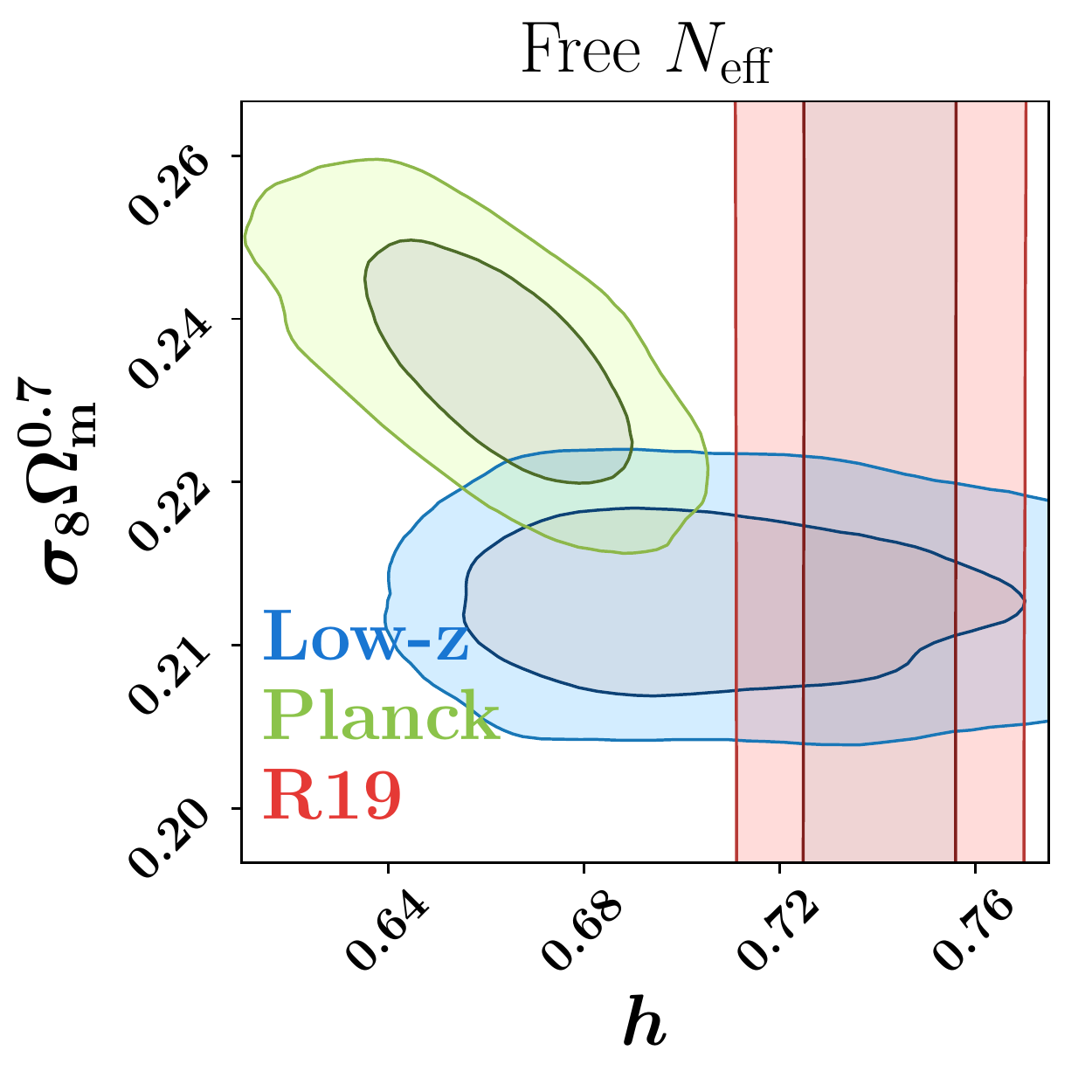}
    \includegraphics[width=0.32\textwidth]{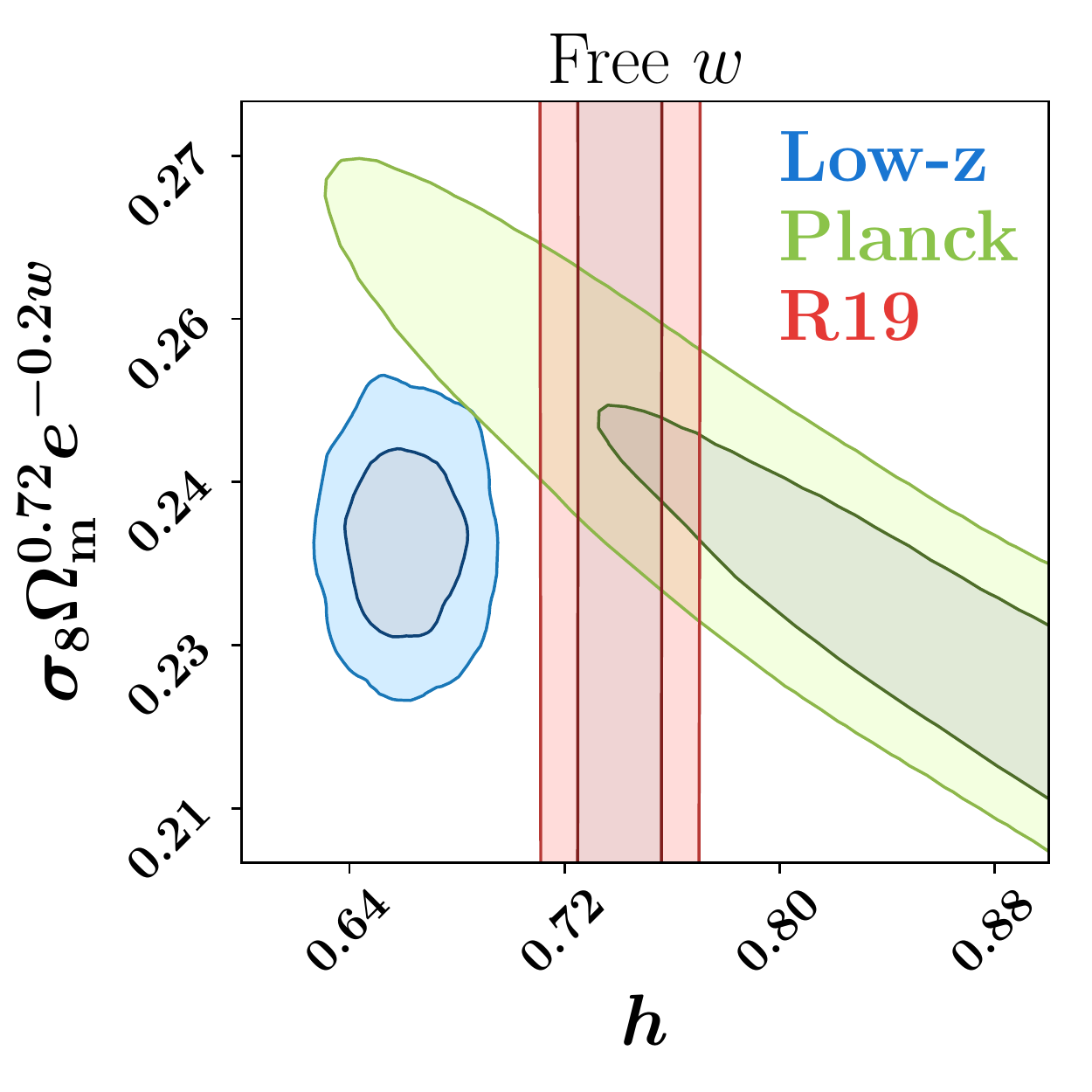}
    \includegraphics[width=0.32\textwidth]{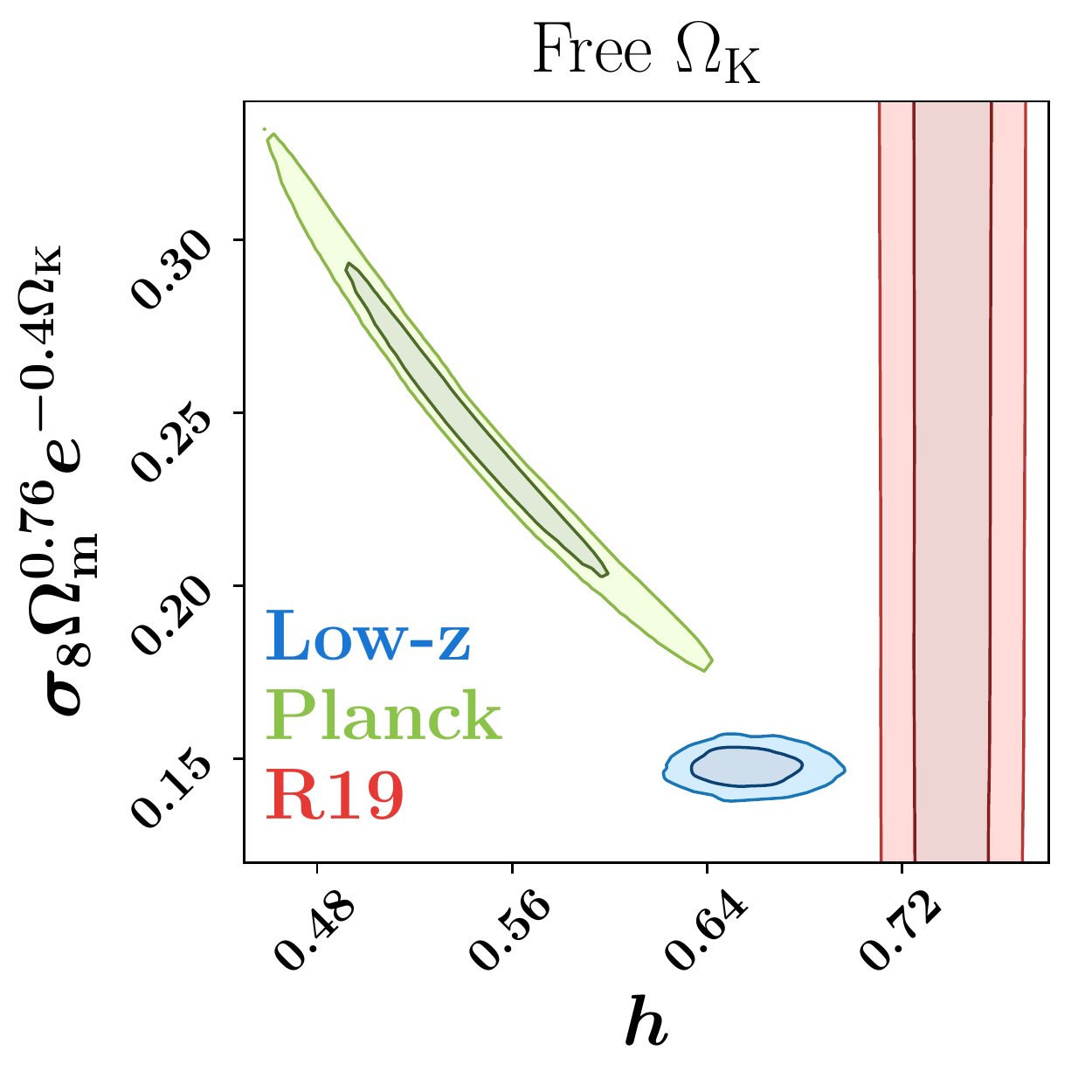}
\end{center}
\caption{\lowz, \planck, and R19 constraints in the relevant 2D parameter subspaces under different extensions -- free $N_\mathrm{eff}$ (left), free $w$ (center), and free $\Omega_\mathrm{K}$ (right) -- of the flat $\Lambda$CDM cosmological model. Note that the axis ranges have been increased from those for Fig. \ref{fig:tension2}.}
\label{fig:tension3}
\end{figure*}


Figure~\ref{fig:tension2} summarizes the different tensions identified in our analysis.  The figure shows the posterior distribution from each of the three data sets considered here in the $e_2$--$h$ ($\etwo$--$h$) plane.  We do not showcase the first eigenvector $e_1$ ($\eone$) because the \lowz\ and \planck\ results are in good agreement along this direction.  

Along the $h$ direction, both \planck\ and \lowz\ are in tension with R19. These two measurements share a common theoretical model for the sound horizon scale $r_s$, but are otherwise fully independent.  That is, the \lowz\ measurement constitutes {\it a new, completely independent calibration of the sound horizon scale}.  The excellent agreement between the \planck\ and \lowz\ measurements of $h$ suggests that the evolution of the angular diameter distance in the flat $\Lambda$CDM model is correct, though a conspiracy of cancelling errors at two epochs is still a possibility.  In absence of such a conspiracy, reconciling these two experiments with R19 will require a modification of the early Universe physics capable of impacting the sound horizon scale. Recently, \citet{ayloretal19} reached a similar conclusions from an in-depth analysis of the acoustic features in the CMB, thereby strengthening our conclusion: new physics that can relieve the tension between R19 and both of the sound horizon based results -- \planck\ \underline{and} \lowz\ -- must reside in the high redshift Universe.

Turning our attention to the $e_2$ ($\etwo$) direction in Figure~\ref{fig:tension2}, it is readily apparent that the \planck\ and \lowz\ data sets are in mild tension with each other.  Unlike the $H_0$ tension, the difference between the \planck\ and \lowz\ data sets points towards new physics in the late-time Universe, likely associated with the current phase of accelerated expansion.  Specifically, the $\etwo$ parameter combination can be thought of as characterizing the ``lensing strength'' of the late-time Universe, itself determined by the amplitude of matter inhomogeneities ($\sigma_8$) and the amount of matter in the late-time Universe ($\Omega_{\rm m}$). This lensing strength can be predicted from the amplitude of inhomogoneities and matter density inferred from the early Universe as measured by \planck. 
That is, the tension between \planck\ and \lowz\ reflects a failure of the flat $\Lambda$CDM model to correctly extrapolate from the high redshift measurements from \planck\ to the low redshift measurements from \lowz.

Taken together, the two distinct tensions highlighted in Figure~\ref{fig:tension2} suggest that, if both tensions persist, reconciling all three data sets may require a new set of physics that impacts both the early and late-time physics of the Universe.  For this reason, many of the usual candidates for extensions of a flat $\Lambda$CDM model are not especially attractive approaches towards simultaneously reconciling all three data sets in Figure~\ref{fig:tension2}. To demonstrate this point, we consider three such extensions: 1) free $N_{\rm eff}$, 2) free $w$, and 3) free $\Omega_\mathrm{K}$.  For each extension, we first obtain the updated \lowz\ eigenmodes and identify the eigenmode $e_2 '$ that exhibits the strongest overlap with the original \lowz\ $e_2$. We then examine constraints from \lowz, \planck, and R19 in the $e_2'$-$h$ plane, in analog to our original comparison in Figure~\ref{fig:tension2}.

We note that \planck\ does not yet provide likelihoods or chains that simultaneously vary the above extension parameters --- more specifically $w$ and $\Omega_\mathrm{K}$ --- together with $\sum m_\nu$. Consequently, we fix $\sum m_\nu=0.06$~eV for the free $w$ and $\Omega_\mathrm{K}$ scenarios to enable comparisons against \planck\ results in these models.

Before presenting the results of each of the above extensions, it is worth working through what we expect to find based on our earlier discussion.  When we allow for an early Universe modification ($N_{\rm eff}$), we allow for freedom in the sound horizon calibration.  This might enable reconciling the sound-horizon based measurements of the Hubble constant with R19, but should do little to abate the tension between \planck\ and \lowz.  Conversely, when we consider modifications of the late-time Universe ($w\neq -1$ or $\Omega_k \neq 0$ ), we may hope that opening up these degrees of freedom will help reconcile \planck\ and \lowz\ estimates of the lensing strength of the low redshift Universe.  However, even if such a reconciliation does happen, we would predict that the tension in the Hubble parameter between R19 and at least one of the two sound horizon based measurements of $H_0$ (\planck\ or \lowz) would remain.
Let's see what happens. 

Figure~\ref{fig:tension3} shows the $e_2'$--$h$ parameter subspace for each of the three extensions we considered.  Freeing $N_\mathrm{eff}$ introduces early universe modifications to the flat $\Lambda$CDM model by changing the radiation density in the early universe. Under this extension, we find
\begin{equation}
    e_2 ' = \sigma_8\Om^{0.7},
\end{equation}
almost identical to the flat $\Lambda$CDM \lowz\ $e_2$. In this model, the \planck--R19 and \lowz--R19 tensions are $3.5\sigma$ and $0.9\sigma$ respectively.  As expected, the tension between the sound horizon measurements and R19 has been reduced, but in the case of \planck\, significant tension remains.  This is because the high-$l$ multipoles in \planck\ effectively constrain $N_{\rm eff}$. Along the new $e_2'$ direction, \lowz\ and \planck\ are in 2.6$\sigma$ tension, similar to the original flat $\Lambda$CDM result.

Next, we consider the $w$CDM model, in which the dark energy equation of state $w$ is allowed to vary.  We obtain 
\begin{equation}
    e_2' = \sigma_8 \Omega_\mathrm{m}^{0.72} e^{-0.2w}.
\end{equation} 
In this extension, the \lowz\ data set exhibits a third robust eigenmode, with a well-measured parameter ($w$) adding an extra dimension. As previously discussed, we first restrict our attention to the $e_2'$ mode to see if the tension along this direction has been alleviated.

As shown in Figure~\ref{fig:tension3}, this late-time Universe modification helps reconcile the tension between the \planck\ and \lowz\ experiments along the $e_2'$ direction. Along the $h$ direction, the tension between the two experiments and R19 respectively changes from 4.0$\sigma$ (\planck) and 3.7$\sigma$ (\lowz) to 1.5$\sigma$ and 3.9$\sigma$, demonstrating that \lowz\ remains in tension with R19. This reveals a subtle difference in the nature of sound horizon calibrations between \planck\ and \lowz. Both methods require calculations of angular diameter distances, which in turn depends on $w$. However, while \planck\ has little constraining power on $w$, \lowz\ tightly constrains $w$ and thus allows much less freedom in the derived angular diameter distances. More specifically, the \textit{Pantheon} data holds the shape of the angular diameter distance as a function of redshift fixed. As a result, the \lowz\ constraints on $h$ remains tight and in significant tension against R19, implying that a free $w$ fails to reconcile the three-way tension highlighted in Figure~\ref{fig:tension3}.

\begin{figure}
\begin{center}
    \includegraphics[width=\columnwidth]{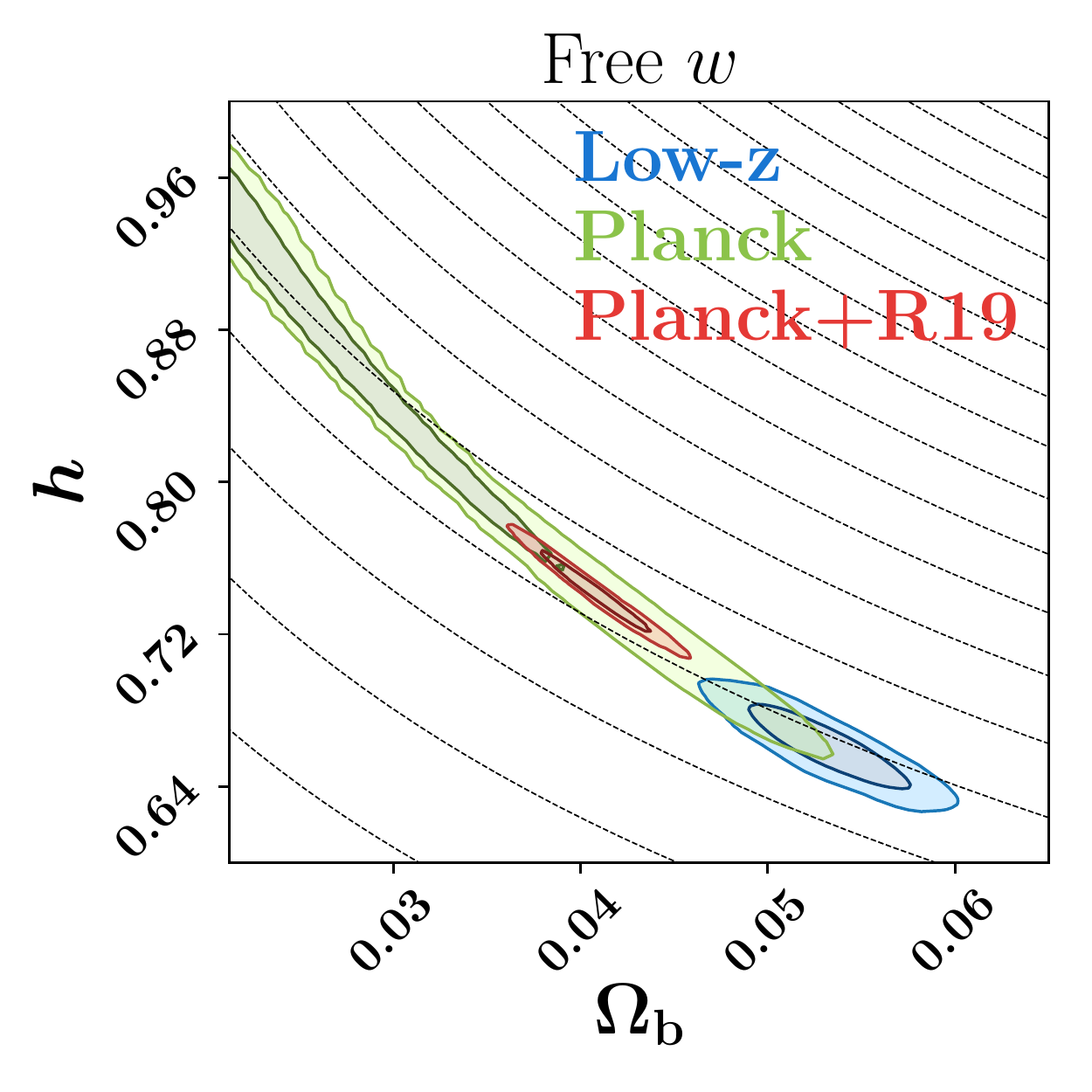}
\end{center}
\caption{Comparison of \lowz, \planck, and \planck+R19 constraints in the $(\Omega_{\mathrm{b}}, h)$ plane. While the \lowz\ (blue) and the \planck+R19 (red) contours are well separated in this plane, they fall along the degeneracy direction of $e_1' = \Ob h^{2.55}$ (shown here by the dashed lines perpendicular to $e_1'$) and thus the experiments exhibit no tension when utilizing our proposed metric.}
\label{fig:tension4}
\end{figure}

It is worth pausing here to consider a hypothetical scenario in which we first studied \planck\ and R19 within the context of a $w$CDM model.  In such a scenario, \planck\ and R19 would be found to be consistent, enabling us to combine the two experiments.  We would have then tested for consistency between \planck\ + R19 as experiment A and \lowz\ as experiment B.  Applying our tension metric, we find that despite the difference in the $h$ posterior noted above, the two experiments are consistent with each other, the difference between them being significant only at the $1.1\sigma$ level.  While \planck\ + R19 prefers significantly higher values of $h$ compared to \lowz, it also prefers lower values of $\Ob$. As a result, \planck+R19 is not very different from \lowz\ along the $w$CDM eigenmode $e_1' = \Ob h^{2.55}$, and no tension is detected despite the clear difference in the posteriors for $h$. This ``failure mode'' of our tension metric is illustrated in Figure \ref{fig:tension4}, and clearly demonstrates that the tension metric considered here is not generically optimal.

Finally, allowing curvature to vary yields
\begin{equation}
    e_2' = \sigma_8 \Omega_\mathrm{m}^{0.76} e^{-0.4\Omega_\mathrm{K}}.
\end{equation}
We see that freeing this degree of freedom does not help reconcile \planck\ with \lowz, and it also fails to reconcile \planck\ with R19.  It is curious that \planck\ is now in strong tension with both \lowz\ and R19.  As discussed in \cite{Planck2018}, this is due to detailed features in the high-$l$ multipoles in \planck, which also lead to anomalous estimates of the lensing amplitude from the high-$l$ multipoles.  Based on the cautionary words in \cite{Planck2018} (see particularly Sect. 7.3), we will not discuss this extension any further.  


In summary, we have argued that the present state of cosmology presents us with not one but two distinct challenges to the supremacy of the flat $\Lambda$CDM model.  On the one hand, cosmological distances inferred from independent calibrations of the sound horizon scale as a standard ruler are in tension with local distance ladder measurements.  If this tension is not ultimately traced to underlying systematics in one or more of the measurements, it  suggests that the physics of the early Universe, particularly around the epoch of recombination, may not be entirely understood.  The day before this paper was submitted to arXiv, the H0LiCOW collaboration published their most recent constraints on $H_0$, finding $H_0=73.3^{+1.7}_{-1.8}\ {\rm km/s/Mpc}$ \citep{h0licow19}.  This constraint, based on strong lensing measurements, is in very good agreement with the R19 measurement, and in tension with the \planck\ and \lowz\ estimates, further strengthening the case for modifications in early Universe physics.

In constrast to the $H_0$ tension, the difference between the predicted low redshift lensing strength of the Universe from \planck\ and that inferred from current cosmic shear experiments suggests that, should these tensions persist, new physics in the late-time Universe may be required, likely related to the current phase of accelerated expansion of the Universe.  In short, simultaneously reconciling cosmic shear experiments, distance ladder measurements, and \planck\ CMB data will likely require modifications of both the early and late-time universe physics (or one modification that impacts cosmological evolution at both times).  Evidently, it is now more urgent than ever to stress test the experimental procedures upon which these results rest, while continuing to decrease the corresponding uncertainties.  If the current tensions persist, we may well be on the verge of not one but two distinct cosmological revolutions. 

\acknowledgements

We would like to thank Vivian Miranda and Masahiro Takada for useful conversations, and for comments on an early version of this draft that significantly improved our presentation.  This work was supported by the DOE grant DE-SC0015975.  ER also acknowledges support by the Cottrell Scholar program of the Research Corporation for Science Advancement.

\bibliography{database.bib}

\end{document}